\documentclass[12pt]{iopart}

\usepackage{epsfig}

\begin{document}

\title{Effect of Zn doping on magnetic order and superconductivity
in LaFeAsO}

\author{Yuke Li, Xiao Lin, Qian Tao, Cao Wang, Tong Zhou, Linjun Li, Qingbo Wang, Mi He, Guanghan Cao\footnote[1]{Electronic address: ghcao@zju.edu.cn}, and Zhu'an Xu\footnote[2]{Electronic address: zhuan@zju.edu.cn}}

\address{Department of Physics, Zhejiang University, Hangzhou 310027, China}

\begin{abstract}
We report Zn-doping effect in the parent and F-doped LaFeAsO
oxy-arsenides. Slight Zn doping in LaFe$_{1-x}$Zn$_{x}$AsO
drastically suppresses the resistivity anomaly around 150 K
associated with the antiferromagnetic (AFM) spin density wave
(SDW) in the parent compound. The measurements of magnetic
susceptibility and thermopower confirm further the effect of Zn
doping on AFM order. Meanwhile Zn doping does not affect or even
enhances the $T_c$ of LaFe$_{1-x}$Zn$_{x}$AsO$_{0.9}$F$_{0.1}$, in
contrast to the effect of Zn doping in high-$T_c$ cuprates. We
found that the solubility of Zn content ($x$) is limited to less
than 0.1 in both systems and further Zn doping (i.e., $x$ $\geq$
0.1) causes phase separation. Our study clearly indicates that the
non-magnetic impurity of Zn$^{2+}$ ions doped in the Fe$_2$As$_2$
layers affects selectively the AFM order, and superconductivity
remains robust against the Zn doping in the F-doped
superconductors.

\end{abstract}


\pacs{74.70.Dd, 74.62.Dh, 74.25.Fy, 74.25.Ha}

\maketitle

\section{Introduction}

Since the discovery of superconductivity at 26 K in 1111 phase
LaFeAsO$_{1-x}$F$_{x}$\cite{Kamihara08}, $T_c$ has been raised up
to 56 K\cite{Chen-Sm,Chen-Ce,Ren-Sm,Wang} quickly in a series of
doped oxy-arsenides. Although a BCS-like superconducting gap has
been observed by the studies of Andreev spectroscopy\cite{TYChen}
and angle-resolved photoemission spectroscopy (ARPES)\cite{ARPES,
HDing}, theoretical calculations have excluded conventional
pairing mechanism based on electron-phonon
interaction\cite{Boeri}. Thus various theoretical models have been
proposed \cite{Mazin, DaiX,Hirsch,Li,ZhangSC,Baskaran,WenXG,Si}.
The evolvement of electronic state with charge carrier doping
indicates that superconductivity occurs on the verge of AFM
ordering state \cite{phasedia}, suggesting the similarities with
high-$T_c$ cuprates.

Elemental substitution is a very useful approach to explore new
superconducting materials and to elucidate the intrinsic factors
which determine $T_c$ in unconventional superconductors. In the
case of high-$T_c$ cuprates, the partial substitution of Cu with
other 3$d$ elements such as Ni and Zn in CuO$_2$ planes destroys
severely superconductivity\cite{Xiao,Tarascon}. However, the band
structure calculations\cite{Singh,Lu,wnl} and theoretical
analysis\cite{Tesanovic} reveal the itinerant character of Fe 3$d$
electrons in the iron-based oxy-arsenides, in comparison with the
localized nature of Cu 3$d$ electrons in cuprates. Therefore
different behaviors could be expected when Fe is partially
substituted by other 3$d$ transition metal elements. Indeed
superconductivity has been observed in both 1111 phase and 122
phase by doping magnetic elements Co\cite{CaoCo,Sefat} or
Ni\cite{CaoNi}. Although previous report has discovered that
LaCoAsO is a ferromagnetic metal with a Curie temperature $T_c$ =
66 K\cite{Yanagi}, which suggests that Co$^{2+}$ ions should be
magnetic in this structure, we found that the Co 3$d$ electrons
are basically itinerant in LnFe$_{1-x}$Co$_x$AsO system
(Ln:lanthanides, $x$ $\leq$ 0.3)\cite{CaoCo}. These studies have
revealed that there are essentially different doping mechanisms
between iron-based arsenides and cuprates, and suggested
fundamental difference between the two classes of high-temperature
superconductivity. The effect of non-magnetic Zn doping in the
iron-based arsenides has not been reported yet. In high-$T_c$
cuprates, Zn doping not only suppresses superconductivity severely
but also induces local magnetic moments and other anomalous
properties\cite{Alloul,Kakurai,Julien}. Therefore it is a very
interesting issue to address.

In this paper we report the investigation on the doping effect of
the non-magnetic element Zn in the parent and F-doped LaFeAsO
systems. We found that even slight Zn doping in
LaFe$_{1-x}$Zn$_{x}$AsO drastically suppresses the resistivity
anomaly around 150 K which is associated with formation of
antiferromagnetic (AFM) spin density wave (SDW) in the undoped
parent compound\cite{Dai}. Measurements of magnetic susceptibility
and thermopower provide further evidence for the suppression of
SDW order. Meanwhile slight Zn doping has little effect on
superconductivity in LaFe$_{1-x}$Zn$_{x}$AsO$_{0.9}$F$_{0.1}$, in
contrast to the effect of Zn doping in high-$T_c$ cuprates. Our
result implies that the Zn doping selectively destroys the SDW
order rather than Cooper pairs.

\section{Experimental}

The polycrystalline LaFe$_{1-x}$Zn$_{x}$AsO and
LaFe$_{1-x}$Zn$_{x}$AsO$_{0.9}$F$_{0.1}$ samples were synthesized
by solid state reaction in vacuum using powders of LaAs,
La$_{2}$O$_{3}$, FeAs, Fe$_{2}$As, LaF$_3$, and ZnO. LaAs was
presynthesized by reacting stoichiometric La pieces and As powders
in evacuated quartz tubes at 1223 K for 24 hours. FeAs and
Fe$_{2}$As were prepared by reacting stoichiometric Fe powders and
As powders at 873 K for 10 hours, respectively. The powders of
these intermediate materials were weighed according to the
stoichiometric ratio of LaFe$_{1-x}$Zn$_{x}$AsO ($x$=0, 0.02,
0.05, 0.10, and 0.15), and
LaFe$_{1-x}$Zn$_{x}$AsO$_{0.9}$F$_{0.1}$ ($x$=0, 0.02, 0.05, 0.1,
and 0.15) respectively, and then thoroughly mixed in an agate
mortar. The mixtures were pressed into pellets under a pressure of
2000 kg/cm$^{2}$. All the processes were operated in a glove box
filled with high-purity argon. The pellets were sealed in
evacuated quartz tubes and heated uniformly at 1433 K for 40
hours.

Powder X-ray diffraction (XRD) was performed at room temperature using a D/Max-rA diffractometer with Cu
K$_{\alpha}$ radiation and a graphite monochromator. The XRD diffractometer system was calibrated using standard
Si powders. Lattice parameters were refined by a least-squares fit using at least 20 XRD peaks.

The electrical resistivity was measured by a standard
four-terminal method. The samples for transport property
measurements were cut into a thin bar with typical size of
4mm$\times$2mm$\times$0.5mm. The size of the contact pads leads to
a total uncertainty in the absolute values of resistivity of ¡À10
\%. Thermopower was measured by using a steady-state technique.
The temperature dependence of d.c. magnetic susceptibility was
measured on a Quantum Design magentic property measurement system
(MPMS-5). The applied magnetic field was 1000 Oe for the
non-superconducting LaFe$_{1-x}$Zn$_{x}$AsO, and 10 Oe for the
superconducting LaFe$_{1-x}$Zn$_{x}$AsO$_{0.9}$F$_{0.1}$. The
superconducting samples used for magnetic measurements were made
in roughly cubic shape so that a shape demagnetizing factor ($N$)
of about 1/3 was used to calculate the demagnetizing effect in the
superconducting state.

\section{Results and discussion}
\subsection{Effect of Zn doping in LaFe$_{1-x}$Zn$_{x}$AsO}

Figure 1 shows the XRD patterns of LaFe$_{1-x}$Zn$_{x}$AsO samples
and the variations of the lattice constants with Zn content. For
the parent compound and slightly Zn doped compounds ($x$ $\leq$
0.05), the XRD peaks can be well indexed based on a tetragonal
cell with the space group of P$4/nmm$, which indicates that the
samples are essentially single phase. With further increasing Zn
content, peak splitting can be observed for the peaks (102),
(112), and (200), which indicates that there is a possible phase
separation for $x$ $\geq$ 0.10. Since the lattice constants of
LaZnAsO are quite larger than those of LaFeAsO\cite{RZnAsO}, a
phase separation of LaZnAsO from the uniform
LaFe$_{1-x}$Zn$_{x}$AsO phase could occur. From the Fig.1 (b), it
can be seen that the $a$-axis increases slightly with $x$, but the
$c$-axis shrinks slightly. The cell volume decreases first and
then increases slightly for $x$ $>$ 0.05. For larger Zn content,
the change in the lattice constants $a$ and $c$ becomes saturated,
which is in agreement with the phase separation as $x$ $\geq$ 0.1.

Figure 2 shows the temperature dependence of resistivity for the
LaFe$_{1-x}$Zn$_{x}$AsO samples. The resistivity of the undoped
parent compound shows a drop around $T^*$ of 150 K, where $T^*$ is
defined as the onset point of this anomaly. As shown in the inset,
the anomaly becomes more clear in the plot of d$\rho$/d$T$ versus
$T$. Such an anomalous drop in the resistivity is associated with
the structural transition and/or formation of antiferromagnetic
(AFM) spin density wave (SDW)\cite{Dai}. Even with addition of
only 2\% Zn content ($x$ =0.02), the anomaly becomes less
significant and moves to a lower temperature ($T^*$ $\sim$ 140 K),
and the resistivity itself becomes semiconductor-like. As $x$
increases to 0.05, this anomaly almost disappears. Only a very
tiny kink around 120 K can be observed in the curve of
d$\rho$/d$T$ versus $T$ for $x$ = 0.05, which might result from
the residual SDW order. On the other hand, in contrast to the case
of Co doping on Fe site\cite{CaoCo,Sefat}, no superconductivity in
the Zn doped LaFe$_{1-x}$Zn$_{x}$AsO samples is observed for
temperature down to 2 K. Instead, the resistivity is
semiconductor-like for all the Zn doped samples. Actually, the
resistivity increases roughly logarithmically with decreasing
temperature at low temperatures instead of usual thermally
activated conductivity in typical semiconductors. It should also
be noted that the resistivity value increases gradually with
increasing Zn content, suggesting the effect of disorder caused by
Zn doping. For $x$ $\geq$ 0.10, there is a large increase in the
resistivity, consistent with the fact that the
LaFe$_{1-x}$Zn$_{x}$AsO samples become phase separated for $x$
$\geq$ 0.10 as shown by the XRD patterns.

Figure 3 shows the temperature dependence of magnetic
susceptibility for the LaFe$_{1-x}$Zn$_{x}$AsO samples. A clear
drop in susceptibility related to SDW order can be observed around
150 K for the parent compound, which is consistent with previous
reports\cite{Kamihara08,Dai}. With increasing Zn content, the
susceptibility increases gradually, and the linear temperature
dependence of susceptibility for the temperature above the SDW
transition, which could result from the fluctuations of "SDW
moments" of Fe$^{2+}$ ions\cite{ZhangGM}, becomes less and less
significant. The Curie-like upturn observed at low temperatures
could be mainly due to an extrinsic origin (such as defects and
trace impurities). For $x$ = 0.02, the drop in susceptibility can
still be observed around 137 K, consistent with the resistivity
result. However, no anomalous change in susceptibility can be
observed for $x$ $\geq$ 0.05. The magnetic data show that SDW
order is very sensitive to Zn doping.

In order to confirm further the suppression of SDW order by Zn
doping, the measurement of thermopower ($S$) was also performed.
As demonstrated by several previous reports on the novel
superconductors like NbSe$_2$ and Sr$_2$RuO$_4$, the thermopower
is very sensitive to subtle changes in the electronic
structure\cite{Bel, XuNst}. Figure 4 shows the temperature
dependence of thermopower for the LaFe$_{1-x}$Zn$_{x}$AsO samples.
A remarkable increase in $S$ occurs around the structural and
magnetic transitions. Similar behavior has been observed in the
previous reports\cite{Guire,LiLJ}. This anomaly in $S$ could
result from opening of a pseudo-gap in Fermi surface. Optical
spectra indeed revealed the evidence for pseudo-gap opening below
structrual/SDW transitions\cite{wnl}. Because of the multi-band
nature, a quantitative analysis of the thermopower data is quite
difficult. However, the anomalous change in $S$ can be regarded as
an experimental evidence of the SDW transition. It can be seen
from Fig. 4 that the anomaly in $S$ becomes less significant and
moves to a lower temperature for $x$ = 0.02, still observable for
$x$ = 0.05, but almost disappears for $x$ $\geq$ 0.10. The
thermopower data confirm further that the SDW order is severely
suppressed by Zn doping, essentially consistent with the
resistivity and susceptibility data. The effect of Zn doping on
the SDW order is striking, which implies that the magnetism in the
Fe$_2$As$_2$ layers can be affected severely by addition of
non-magnetic Zn$^{2+}$ ions. However, as shown below, there is
almost no influence of Zn doping on superconductivity in the
F-doped LaFe$_{1-x}$Zn$_{x}$AsO$_{0.9}$F$_{0.1}$.

\subsection{Superconductivity in Zn doped LaFe$_{1-x}$Zn$_{x}$AsO$_{0.9}$F$_{0.1}$}

Figure 5(a) shows the XRD patterns of
LaFe$_{1-x}$Zn$_{x}$AsO$_{0.9}$F$_{0.1}$ samples and Figure 5(b)
shows the variations of the lattice constants with Zn content. For
small Zn doping level, all the peaks can be well indexed based on
a tetragonal cell with the space group of P$4/nmm$. Obviously F
doping causes remarkably shrinkage of both $a$ and $c$-axes,
similar to the previous report\cite{Kamihara08}. With the addition
of Zn in the F doped samples, the $a$-axis decreases slightly ,
while the $c$-axis increases. The change in the $c$-axis becomes
saturated for $x$ $\geq$ 0.05, indicating that the samples could
become phase separated. A few tiny foreign XRD peaks corresponding
to the impurity phases LaZnAsO, La$_2$O$_3$, and FeAs appear for
large $x$ (indicated by symbol $\sharp$, $\star$, and $\$$,
respectively). As in the case of LaFe$_{1-x}$Zn$_{x}$AsO, there is
a limit of Zn solubility on the Fe-site and superabundant Zn
doping leads to phase separation. The solubility of Zn doping in
the F-doped LaFeAsO$_{0.9}$F$_{0.1}$ is even smaller compared to
the parent compound.

Figure 6 shows the temperature dependence of resistivity for the
LaFe$_{1-x}$Zn$_{x}$AsO$_{0.9}$F$_{0.1}$ ($x$ = 0, 0.02, 0.05,
0.1, and 0.15) samples, and the inset shows the temperature
dependence of d.c. magnetic susceptibility measured under $H$ of
10 Oe. Without Zn doping, the 10\% F doping (almost at the optimal
doping level) induces superconductivity at $T_{c}$$^{onset}$ of
26.8 K and $T_c$$^{mid}$ of 23.0 K, where $T_{c}$$^{onset}$ and
$T_c$$^{mid}$ are defined as the onset point and midpoint
temperature in the resistive transition respectively, consistent
with the previous report\cite{Kamihara08}. To our surprise, the
temperature dependence of resistivity becomes even more metallic
for the Zn doped samples. Meanwhile $T_c$$^{mid}$ slightly
increases to 25.5 K and 25.4 K for $x$ = 0.02 and 0.05
respectively. The volume fraction of superconducting magnetic
shielding is nearly 100 \% for the samples with $x$ = 0, 0.02 and
0.05 estimated from the magnetic susceptibility. The demagnetizing
factor $N$ is taken into account in our estimation, but the volume
fraction could be still a little over-estimated due to other
factors. The sharp transitions in the magnetic susceptibility
suggest bulk superconductivity and high homogeneity of the
Zn-doped samples for $x$ $\leq$ 0.05. For $x$ $\geq$ 0.10, there
is a large increase in the normal state resistivity, and the
superconducting transition becomes broad due to the phase
separation. Meanwhile, the volume fraction of superconducting
magnetic shielding also decreases quickly for $x$ $\geq$ 0.05,
indicating the existence of non-superconducting impurity phases.
To our surprise, Zn doping does not suppress superconductivity in
the single phase F-doped samples, and it even enhances $T_c$
instead.

The temperature dependence of thermopower for the
LaFe$_{1-x}$Zn$_{x}$AsO$_{0.9}$F$_{0.1}$ samples is shown in
Figure 7. The thermopower drops to zero quickly below $T_c$.
Similar to the resistive and magnetic superconducting transitions,
the transition in $S$ for $x$ $\leq$ 0.05 is very sharp, but it
becomes broad for $x$ =0.1 due to the phase separation. The
absolute value of thermopower, $|S|$, is much larger than that of
the parent compound LaFeAsO. There seems to exist a close
correlation between the absolute value of thermopower and $T_c$,
namely, the higher $T_c$ corresponds to larger normal state $|S|$,
as proposed in the previous reports\cite{CaoCo}. There should be a
relationship between the origin of the enhanced thermopower and
the superconducting mechanism.

As is well known, Zn doping on the CuO$_2$ planes causes radical
suppression of superconductivity in the high-$T_c$
cuprates\cite{Xiao,Tarascon}, and interpretations in the frame of
pair-breaking or decrease in the superfluid density were
proposed\cite{XuZA}. In contrast to the case of cuprates, Zn
doping almost has no influence on superconductivity in the
Fe-based oxy-arsenides. This result reveals a significant
difference between high-$T_c$ ferro-arsenides and high-$T_c$
cuprates. In the latter case, any substitution of Cu by other
elements in the CuO$_2$ layers can lead to sharp suppression of
superconductivity. Meanwhile in the ferro-arsenides
superconductivity can survive when the Fe element in the
conducting layer is partially substituted by other 3$d$ transition
elements like Co, Ni, and Zn no mater whether the substitution is
magnetic or non-magnetic. However, there is still an essential
difference between Co doping and Zn doping. That is, the Zn doping
itself cannot induce superconductivity in LaFeAsO system. The
calculation of DOS versus energy for LaO$M$As ($M$ = Mn, Fe, Co
and Ni) by Xu et al.\cite{XuG} shows that the main feature of
total DOS remains unchanged in these systems, except that Fermi
level shifts toward the top of valence band with band filling
(adding electrons) one by one from $M$ = Mn, Fe, Co to Ni. This
study indicates that the 3$d$ electrons of $M$ = Mn, Fe, Co and Ni
are essentially itinerant. The electron configuration of Zn$^{2+}$
ion is 3$d$$^{10}$, very different from that of Mn, Fe, Co or Ni.
A recent band structure calculation of LaZnAsO shows that the
fully occupied Zn 3$d$ states are located about -7 eV to the Fermi
level\cite{Bannikov}, namely, the 3$d$ electrons of Zn in this
structure are localized. Thus the partial substitution of Fe by Zn
is not expected to add more itinerant electrons into FeAs layers,
and it causes mainly disorder instead, which should account for
the difference between the effect of Zn doping and that of Co or
Ni doping in this structure.

Whereas the anomaly in the resistivity is completely suppressed in
the parent compound, the superconductivity in the F-doped system
remains almost unperturbed by Zn doping, which reveals that doped
Zn$^{2+}$ ions affects selectively AFM order and meanwhile the
disorder caused by Zn doping has little effect on the
superconducting electron pairing. Neutron studies have revealed
that the parent compound LaFeAsO is a long-range ordered
antiferromagnet with a simple stripe-type AFM structure within the
plane at low temperatures\cite{Dai}. Such a stripe-type AFM
long-range order can be easily destroyed by disorder. The
substitution of magnetic Fe$^{2+}$ ions with non-magnetic
Zn$^{2+}$ ions destroys the long-range AFM order and short-range
AFM fluctuations could still remains. In the F-doped
LaFe$_{1-x}$Zn$_{x}$AsO$_{0.9}$F$_{0.1}$ superconductor, the
stripe-type long range AFM order exists no longer, and thus the
main effect of Zn doping is to induce the disorder in the FeAs
layers. All the results of Co, Ni or Zn doped superconducting
systems show that high-$T_c$ superconductivity is quite robust
against the disorder in the conducting FeAs layer, which may be
due to the quite three-dimensional superconductivity of the
iron-based arsenide superconductors\cite{Yuan}.

\section{Conclusion}

In summary, we have investigated the effect of the non-magnetic Zn
doping on AFM order and superconductivity in the parent and
F-doped LaFeAsO. A striking discovery is that Zn doping suppresses
severely the AFM order without occurrence of superconductivity in
the parent compound. Meanwhile superconductivity remains almost
unperturbed by Zn doping in the F-doped LaFeAsO systems. We also
found that the solubility of Zn doping on Fe-site is limited to
$x$ = 0.1. Our results may shed light on the mechanism of
superconductivity and the relationship between AFM (SDW) order and
superconductivity.

\section*{Acknowledgments}
This work is supported by the Natural Science Foundation of China,
National Basic Research Program of China (No.2006CB601003 and
2007CB925001) and the PCSIRT of the Ministry of Education of China
(IRT0754).

\pagebreak[4]

\begin{figure}
\includegraphics[width=10cm]{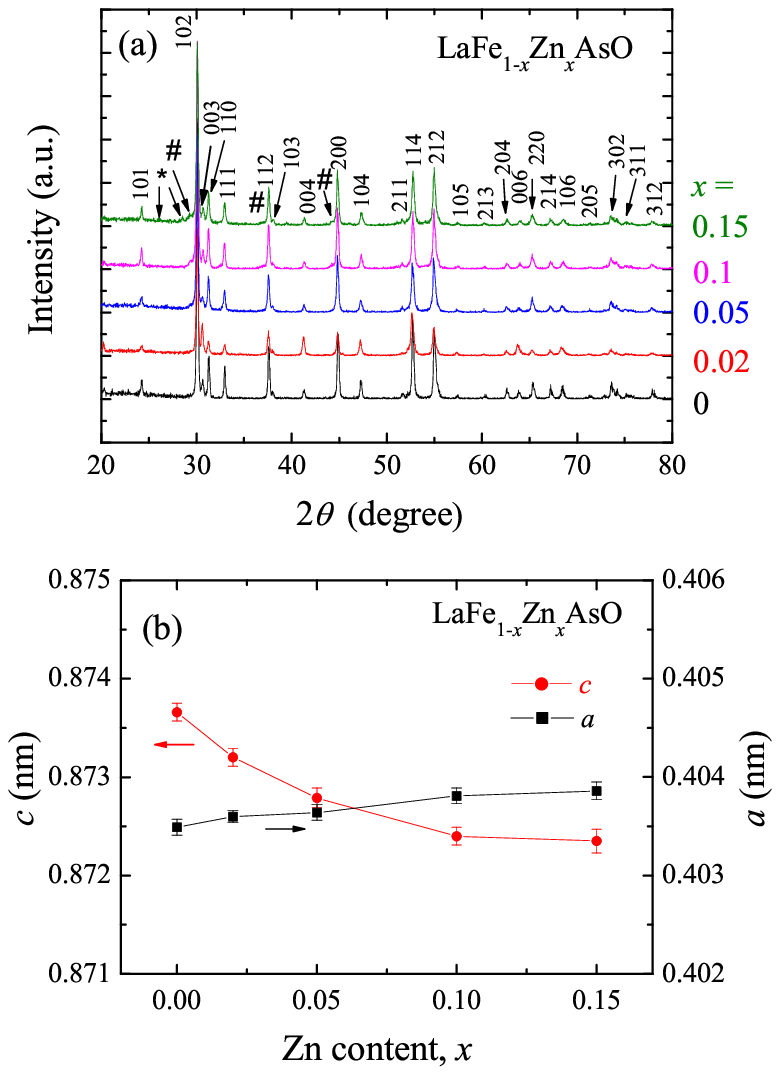}
\caption{\label{Fig.1} (Color Online) (a) Powder X-ray diffraction
patterns of LaFe$_{1-x}$Zn$_{x}$AsO ($x$ = 0, 0.02, 0.05, 0.1 and
0.15 ) samples. The symbols $\sharp$ and $\star$ denote the
impurity phases LaZnAsO and La$_2$O$_3$ respectively. (b)
Variations of lattice parameters as a function of Zn content.}
\end{figure}

\begin{figure}
\includegraphics[width=10cm]{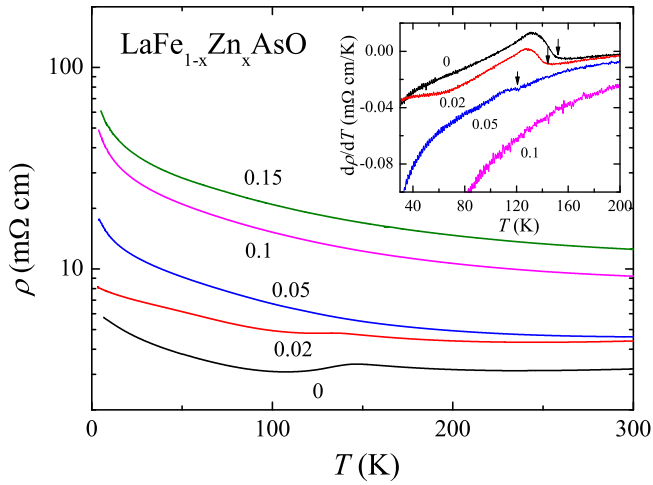}
\caption{\label{Fig.2} (Color Online) Temperature dependence of
resistivity ($\rho$) for LaFe$_{1-x}$Zn$_{x}$AsO samples. The
inset shows the derivative of resistivity versus temperature. The
arrows indicate the anomaly in resistivity which could be
associated to the structural and/or magnetic transition.}
\end{figure}

\begin{figure}
\includegraphics[width=10cm]{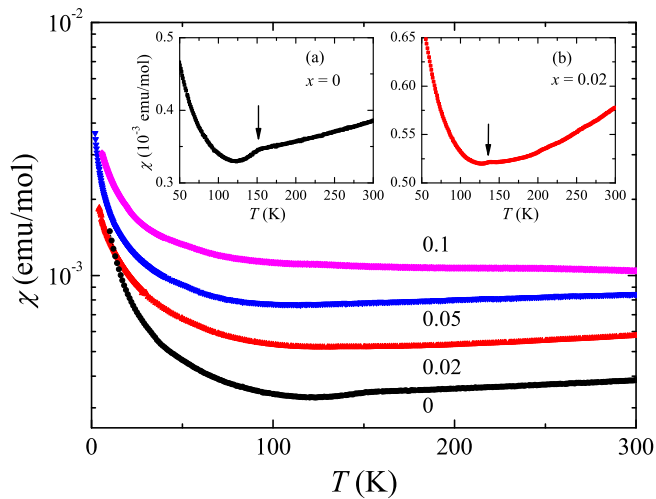}
\caption{\label{Fig.3} (Color Online) Temperature dependence of
magnetic susceptibility ($\chi$) of LaFe$_{1-x}$Zn$_{x}$AsO
samples. The insets show the enlarged plots for $x$ = 0 and 0.02.
The arrows indicate the anomaly in susceptibility related to
structural and/or magnetic transition.}
\end{figure}

\begin{figure}
\includegraphics[width=10cm]{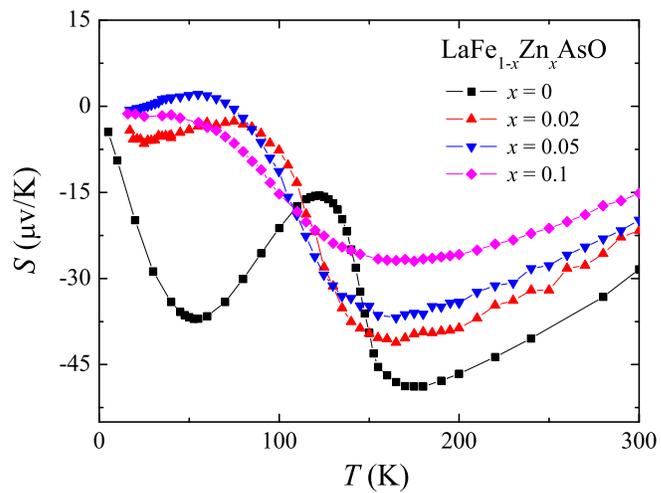}
\caption{\label{Fig.4} (Color Online) Temperature dependence of
thermopower ($S$) for LaFe$_{1-x}$Zn$_{x}$AsO samples.}
\end{figure}

\begin{figure}
\includegraphics[width=10cm]{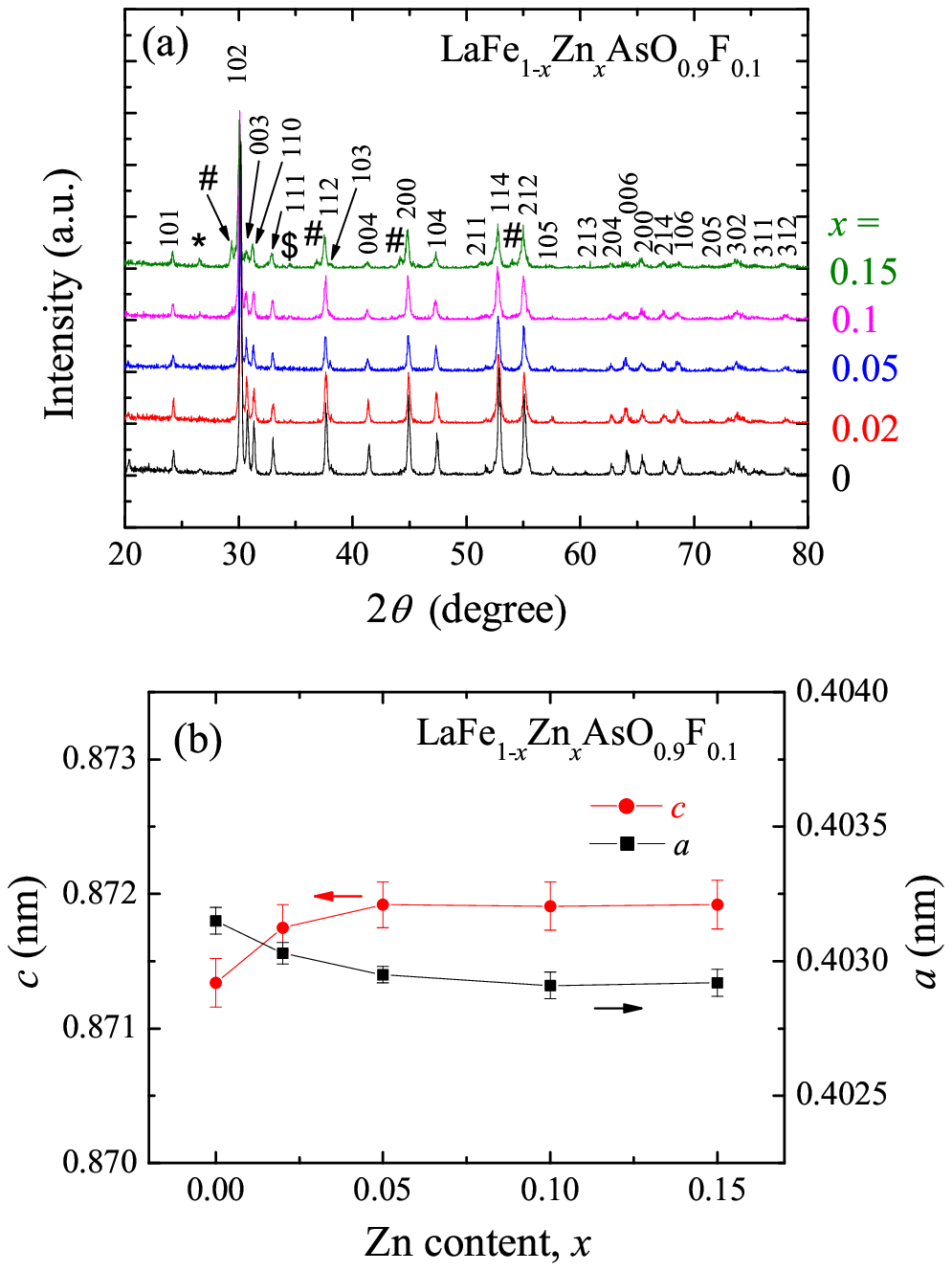}
\caption{\label{Fig.5} (Color Online) (a) Powder X-ray diffraction
patterns of LaFe$_{1-x}$Zn$_{x}$AsO$_{0.9}$F$_{0.1}$ ($x$ = 0,
0.02, 0.05, 0.1, and 0.15) samples. The symbols $\sharp$, $\star$,
and $\$$ denote the impurity phases LaZnAsO, La$_2$O$_3$, and
FeAs, respectively. (b) Lattice parameters as a function of Zn
content.}
\end{figure}

\begin{figure}
\includegraphics[width=10cm]{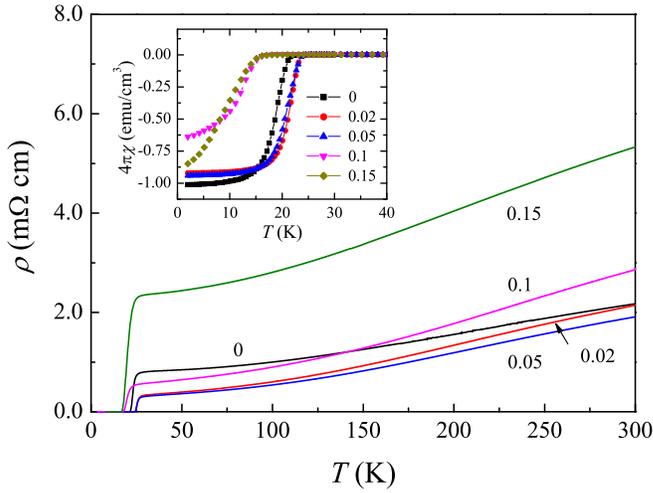}
\caption{\label{Fig.6} (Color Online) Temperature dependence of
resistivity ($\rho$) of LaFe$_{1-x}$Zn$_{x}$AsO$_{0.9}$F$_{0.1}$
samples. Inset: the d.c. magnetic susceptibility ($\chi$) versus
temperature measured under magnetic field of 10 Oe.}
\end{figure}

\begin{figure}
\includegraphics[width=10cm]{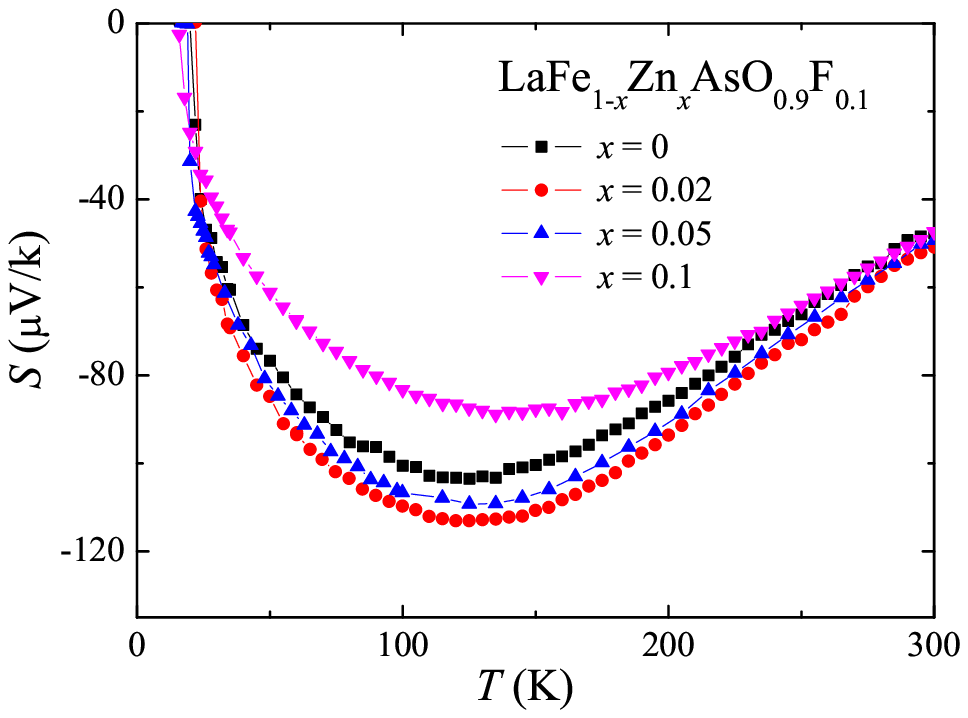}
\caption{\label{Fig.7} (Color Online) Temperature dependence of
thermopower ($S$) for LaFe$_{1-x}$Zn$_{x}$AsO$_{0.9}$F$_{0.1}$
samples.}
\end{figure}


\section*{References}
\begin{thebibliography}{10}

\bibitem{Kamihara08} Kamihara Y, Watanabe T, Hirano M, and Hosono H 2006 {\it J. Am. Chem.
Soc.}  \textbf{130} 3296
\bibitem{Chen-Sm} Chen X H, Wu T, Wu G, Liu R H, Chen H, and Fang D F 2008 {\it
Nature} \textbf{453} 761
\bibitem{Chen-Ce} Chen G F, Li Z, Wu D, Li G, Hu W Z, Dong J, Zheng P, Luo J L, and
Wang N L 2008 {\it Phys. Rev. Lett.} \textbf{100} 247002
\bibitem{Ren-Sm} Ren Z A, Lu W, Yang J, Yi W, Shen X L, Li Z C, Che G C, Dong X L,
Sun L L, Zhou F, and Zhao Z X 2008 {\it Chin. Phys. Lett.}
\textbf{25} 2215
\bibitem{Wang} Wang C, Li L J, Chi S, Zhu Z W, Ren Z, Li Y K, Wang Y T, Lin X, Luo
Y K, Xu X F, Cao G H, and Xu Z A 2008 {\it Europhys. Lett.}
\textbf{83} 67006
\bibitem{TYChen} Chen T Y, Tesanovic Z, Liu R H, Chen X H, and Chien C L  2008 {\it Nature} \textbf{453} 1224
\bibitem{ARPES}  Kondo T, Santander-Syro A F, Copie O, Liu C, Tillman M E, Mun E D, Schmalian J, Bud'ko S L,
Tanatar M A, Canfield P C, and Kaminski A  2008 {\it Phys. Rev.
Lett.} \textbf{101} 147003
\bibitem{HDing} Ding H, Richard P, Nakayama K, Sugawara T, Arakane T, Sekiba Y, Takayama A, Souma S,
Sato T, Takahashi T, Wang Z, Dai X, Fang Z, Chen G F, Luo J L, and
Wang N L 2008 {\it Europhys. Lett.} \textbf{83} 47001
\bibitem{Boeri} Boeri L, Dolgov O V, and Golubov A A 2008 {\it Phys. Rev. Lett.} \textbf{101} 026403
\bibitem{Mazin} Mazin I I, Singh D J, Johannes M D, and Du M H
 2008 {\it Phys. Rev. Lett.} \textbf{101} 057003
\bibitem{DaiX}  Dai X, Fang Z, Zhou Y, and Zhang F C 2008 {\it Phys. Rev. Lett.} \textbf{101} 057008
\bibitem{Hirsch} Marsiglio F and Hirsch J E 2008 {\it Physica C} \textbf{468} 1047
\bibitem{Li} Li T 2008 {\it J. Phys.: Condens. Matter} \textbf{20}
425203

\bibitem{ZhangSC} Raghu S, Qi X L, Liu C X, Scalapino D J, and Zhang S C 2008 {\it Phys. Rev. B} \textbf{77} 220503(R)
\bibitem{Baskaran} Baskaran G 2008 {\it arXiv:} 0804.1341v2
\bibitem{WenXG} Lee P A and Wen X G 2008 {\it arXiv:} 0804.1739v2
\bibitem{Si} Si Q M and Abrahams E 2008 {\it arXiv:} 0804.2480
\bibitem{phasedia}For example, see Zhao J, Huang Q, Cruz C, Li S L, Lynn J W, Chen Y, Green M A,
Chen G F, Li G, Li Z, Luo J L, Wang N L, and Dai P C 2008 {\it
Nature Mater.} \textbf{7} 953
\bibitem{Xiao} Xiao G, Cieplak M Z, Gavrin A, Streitz F H, Bakhshai A, and Chien C L 1988 {\it Phys. Rev. Lett.} \textbf{60} 1446
\bibitem{Tarascon} Tarascon J M, Greene L H, Barboux P, McKinnon W R, Hull G W, Orlando T P, and Delin K A 1987 {\it Phys. Rev. B} \textbf{36} 8393
\bibitem{Singh} Singh D J and Du M H 2008 {\it Phys. Rev. Lett.} \textbf{100} 237003
\bibitem{Lu} Ma F J, Lu Z Y, and Xiang T  2008 {\it Phys Rev. B.} \textbf{78} 224517.

\bibitem{wnl} Dong J, Zhang H J, Xu G, Li Z, Li G, Hu W Z,Wu D,Chen G F, Dai X,Luo J L, Fang Z, and Wang N L 2008 {\it Europhys. Lett.} \textbf{83} 27006
\bibitem{Tesanovic} Cvetkovic V and Tesanovic Z 2009 {\it Europhys. Lett.}
\textbf{85} 37002

\bibitem{CaoCo} Wang C, Li Y K, Zhu Z W, Jiang S, Lin X, Luo Y K, Chi S, Li L J, Ren Z, He M, Chen H, Wang Y T, Tao Q, Cao G H, and Xu Z A
2009 {\it Phys. Rev. B} \textbf{79} 054521
\bibitem{Sefat} Sefat A S, Huq A, McGuire M A, Jin R, Sales B C, and
Mandrus D 2008 {\it Phys. Rev. B} \textbf{78} 104505
\bibitem{CaoNi} Cao G H, Jiang S, Lin X, Wang C, Li Y K, Ren Z, Tao Q, Dai J H, Xu Z A, and Zhang F C
 2008 {\it arXiv:} 0807.4328
\bibitem{Yanagi} Yanagi H, Kawamura R, Kamiya T, Kamihara Y, Hirano M, Nakamura T, Osawa H, and Hosono H 2008 {\it Phys. Rev. B} \textbf{77} 224431
\bibitem{Alloul} Alloul H, Mendels P, Casalta H,  Marucco J F, and Arabski J
1991 {\it Phys. Rev. Lett.} \textbf{67} 3140
\bibitem{Kakurai} Kakurai K, Shamoto S, Kiyokura T, Sato M, Tranquada J M, and Shirane G
 1993 {\it Phys. Rev. B} \textbf{48} 3485
\bibitem{Julien} Julien M H, Merithew R D, Weissman M B, Hess F M, P Spradling, Nowa E R, O'Donnell J, Eckstein J N, Tokura Y, and Tomioka Y
2000 {\it Phys. Rev. Lett.} \textbf{84} 3422
\bibitem{Dai} Cruz C, Huang Q, Lynn J W, Li J Y, Ratcliff W, Zarestky J L, Mook H A,
Chen G F, Luo J L, Wang N L, and Dai P C 2008 {\it Nature}
\textbf{453} 899
\bibitem{RZnAsO} Nientiedt A T and Jeitschko W 1998 {\it Inorg.
Chem.} \textbf{37} 386


\bibitem{ZhangGM} Zhang G M, Su Y H, Lu Z Y, Weng Z Y, Lee D H, and Xiang T 2008 {\it arXiv:} 0809.3874

\bibitem{Bel} Bel R, Behnia K, and Berger H 2003 {\it Phys. Rev. Lett.} \textbf{91}
066602
\bibitem{XuNst} Xu X F, Xu Z A, Liu T J, Fobes D, Mao Z Q, Luo J L, and Liu Y
2008 {\it Phys. Rev. Lett.} \textbf{101} 057002
\bibitem{Guire} Mc Guire M A, Christianson A D, Sefat A S, Sales B C, Lumsden M D, Jin R Y, Payzant E A, and Mandrus D 2008 {\it Phys. Rev. B} \textbf{78}
094517
\bibitem{LiLJ} Li L J, Li Y K, Ren Z, Luo Y K, Lin X, He M, Tao Q, Zhu Z W, Cao G H, and Xu Z A
 2008 {\it Phys. Rev. B} \textbf{78} 132506

\bibitem{XuZA} For example, see Xu Z A, Shen J Q, S R Zhao, Zhang Y J, and Ong C K  2005 {\it Phys. Rev. B} \textbf{72} 144527 and references herein.
\bibitem{XuG} Xu G, Ming W, Yao Y, Dai X, Zhang S C, and Fang Z 2008 {\it Europhys. Lett.} \textbf{82} 67002
\bibitem{Bannikov} Bannikov V V, Shein I R, and Ivanovskii A L 2008 {\it arXiv:}
0810.2606

\bibitem{Yuan} Yuan H Q, Singleton J, Balakirev F F, Baily S A, Chen G F, Luo J L, and Wang N L 2009 {\it Nature} \textbf{457} 565


\end{thebibliography}
\end{document}